\documentclass[12pt,preprint]{aastex}

\def\ifundefined#1{\expandafter\ifx\csname#1\endcsname\relax}

\newif\ifpdf
\ifx\pdfoutput\undefined
   \pdffalse      
\else
   \pdfoutput=1   
   \pdftrue
\fi

\def\la{\mathrel{\hbox{\rlap{\hbox{\lower4pt\hbox{$\sim$}}}\hbox{$<$}}}}
\def\ga{\mathrel{\hbox{\rlap{\hbox{\lower4pt\hbox{$\sim$}}}\hbox{$>$}}}}

\newcommand{\be}{\begin{eqnarray}}
\newcommand{\ee}{\end{eqnarray}}

\ifundefined{ensuremath}\def\ensuremath#1{\relax\ifmmode{#1}}
\else${#1}$\fi\else\relax\fi
\ifundefined{nuc}\def\nuc#1#2{\relax\ifmmode{}^{#1}{\protect\text{#2}}
\else${}^{#1}$#2\fi}\else\relax\fi

\newcommand{\kmps}{\ensuremath{\mbox{km~s}^{-1}}}

\received{} 
\accepted{} 
\journalid{}{} 
\articleid{}{} 
\shortauthors{Baron, E. et~al.}
\shorttitle{SEAM Distance to SN 1999em}

\begin{document}

\title{Type IIP Supernovae as Cosmological Probes: A SEAM Distance to
  SN 1999em}

\author{ E.~Baron,\altaffilmark{1,2,3}\email{baron@nhn.ou.edu} Peter E.~Nugent,
\altaffilmark{2}\email{penugent@lbl.gov} David
Branch,\altaffilmark{1}\email{branch@nhn.ou.edu}  and  Peter
H.~Hauschildt\altaffilmark{4}\email{yeti@hs.uni-hamburg.de}
 }

\altaffiltext{1}{Department of Physics and Astronomy, University of
Oklahoma, 440 West Brooks, Rm.~131, Norman, OK 73019, USA}

\altaffiltext{2}{Computational Research Division, Lawrence Berkeley
  National Laboratory, MS 50F-1650, 1 Cyclotron Rd, Berkeley, CA
  94720-8139 USA}

\altaffiltext{3}{Laboratoire de Physique Nucl\'eaire et de Haute
Energies, CNRS-IN2P3, University of Paris VII, Paris, France}

\altaffiltext{4}{Hamburger Sternwarte, Gojenbergsweg 112,
21029 Hamburg, Germany}

\begin{abstract}

\end{abstract}

\begin{abstract}
Due to their intrinsic brightness, supernovae make excellent
cosmological probes.   We describe the SEAM
method for obtaining distances to Type IIP supernovae (SNe IIP) and
present a distance to SN~1999em for which a Cepheid distance
exists. Our models give results consistent with the Cepheid distance,
even though we have not attempted to tune the underlying
hydrodynamical model, we have simply chosen the best fits. This is in
contradistinction to the expanding photosphere method (EPM) which
yields a distance to SN~1999em that is 50\% smaller than the Cepheid
distance. We emphasize the differences between SEAM and EPM. We show
that the dilution factors used in the EPM analysis were
systematically too small at later epochs. We also show that the EPM
blackbody assumption is suspect. 

Since SNe
IIP are visible to redshifts as high as $z \la 6$, with the
\emph{JWST}, SEAM may be a valuable probe of the early universe.
\end{abstract}

\section{Distances from Supernovae}

A reliable way to determine accurate
distances is a Holy Grail of astronomy and particularly cosmology.  In
order to determine the values of the fundamental cosmological
parameters, an accurate distance indicator visible to high redshift is
required.  Supernovae are extremely bright and hence can be detected
at cosmological distances with modern large telescopes.  Due to their
homogeneity, SNe Ia had long been thought of as as good distance
indicators since they roughly meet the astronomer's definition of a
``standard candle'', that is that the luminosity at peak,
$L_{\mbox{max}}$, is approximately constant. Two Hubble Space
Telescope (\emph{HST}) projects \citep{WFfinal01,parodi00} were
awarded time to use Cepheid variable stars to determine distances to
the Virgo cluster and to determine the Hubble constant to 10\%
accuracy. An additional aim of the program of Sandage and
collaborators \citep{parodi00} was to calibrate the luminosity of SNe
Ia by obtaining Cepheid distances to galaxies which also were the
hosts of SNe~Ia. Distances obtained using Cepheids are considered to
be the among the most reliable in astronomy (purely trigonometric
methods cannot be used at distances in the Hubble flow), but they are
not free of systematic errors and Cepheids are too dim to be observed
at large distances. The reliability of SNe Ia as distance indicators
improved significantly with the realization that the luminosity at
peak was correlated with the width of the light curve \citep{philm15}
and hence that SNe~Ia were correctable candles in much the same way
that Cepheids are \citep{philetal99,goldhetal01,rpk95}. This work and
the development of highly efficient search strategies \citep{perlq097}
sparked two groups to use SNe~Ia to measure the deceleration parameter
and to discover the dark energy \citep{riess_scoop98,perletal99}.

All of the work with SNe~Ia is empirical, based on observed SNe~Ia
template light curves. Another method of determining distances using
supernovae is the ``expanding photosphere method''
\citep[EPM,][]{kkepm,branepm,eastkir89,esk96} a variation of the
Baade-Wesselink method \citep{baadeepm}. The EPM method assumes
that for SNe~IIP, with intact hydrogen envelopes, the spectrum is
not far from that of a blackbody and hence the luminosity is
approximately given by
\[
L = 4\pi\,\zeta^2\,R^2\,\sigma\,T^4
\]
where $R$ is the radius of the photosphere, $T$ is the effective
temperature,  $\sigma$ is the radiation constant, and $\zeta$ is the
``dilution factor'' which takes into account that in a scattering
dominated atmosphere the blackbody is
diluted \citep*{hlw86a,hlw86b,hw87}. The temperature is found
from observed colors, so in fact is a color temperature and not an
effective temperature, the photospheric velocity can be estimated from
observed spectra using the velocities of the weakest lines, 
\[ R = v\, t, \]
the dilution factor is estimated from synthetic spectral models,
and $t$ comes from the light curve and demanding self-consistency.

Both an advantage and disadvantage of EPM is that it primarily requires
photometry. Spectra are only used to determine the photospheric
velocity, colors yield the color temperature, which in turn is used to
determine 
the appropriate dilution factor (from model results).
This method suffers from uncertainties in determining the dilution
factors, the difficulty of knowing which lines to use as velocity
indicators, uncertainties between color temperatures and effective
temperatures, and questions of how to match the photospheric radius
used in the models to determine the dilution factor and the radius of
the line forming region \citep{hamuyepm01,leonard99em02}.  In spite of
this the EPM method was successfully applied to SN~1987A in the
LMC \citep{eastkir89,bran87a} which led to hopes that the EPM method
would lead to accurate distances, independent of other astronomical
calibrators. Recently, the EPM method was applied to the very well
observed SN~IIP
1999em \citep{hamuyepm01,leonard99em02,elmhhamdietal99em03}. All three
groups found a distance of 7.5--8.0~Mpc. 
\citet{leonard99em03} subsequently used \emph{HST} to obtain a
Cepheid distance to the parent galaxy of SN~1999em, NGC 1637, and found
$11.7 \pm 1.0$~Mpc, a value 50\% larger than that obtained with EPM.

With modern detailed NLTE radiative transfer codes, accurate synthetic
spectra of all types of supernovae can be calculated.  The
\textbf{S}pectral-fitting \textbf{E}xpanding \textbf{A}tmosphere
\textbf{M}ethod  \citep[SEAM,][]{b93j3,b94i1,l94d01,mitchetal87a02} was
developed using the generalized stellar atmosphere code
\texttt{PHOENIX} 
\citep[for a review of the code see][]{hbjcam99}. While SEAM is 
similar to EPM in spirit, it avoids the use of dilution factors
and color temperatures. Velocities are determined accurately by
actually fitting synthetic  and observed spectra. The radius is
still determined by the relationship $R = vt$, (which is an excellent
approximation because all supernovae quickly reach homologous
expansion) and the explosion time is found by demanding self consistency.
SEAM uses all the spectral information available in the observed spectra
simultaneously which broadens the base of parameter determination.
Since the spectral energy distribution is known completely from the
calculated synthetic spectra, one may calculate the absolute
magnitude, $M_X$,
in any photometric band $X$, 
\[
M_{X} = -2.5 \log \int_{0}^{\infty} S_{X}(\lambda)\, L_{\lambda}\,
d\lambda + C_X \,
\]
where $S_{X}$ is the response of filter $X$, $L_{\lambda}$ is the
luminosity per unit wavelength, and $C_X$ is the zero point of filter
$X$ determined from standard stars. Then one immediately obtains a
distance modulus $\mu_X$, which is a measure of the distance
\[ \mu_X \equiv m_X - M_X - A_X = 5\log{(d/10\textrm{pc})}, \]
where $m_X$ is the apparent magnitude in band $X$ and $A_X$ is the
extinction due to dust along the line of sight both in the host
galaxy and in our own galaxy. \citet{bsn99em00} found that
the early spectra were quite sensitive to the assumed reddening and
hence determined a value of $E(B-V) =0.1$ for SN~1999em.  The SEAM
method does not need to invoke a blackbody assumption or to calculate
dilution factors.

\section{Results}

We used the above method to calculate the distance to SN~1999em. The
models were taken from Model S15 of \citet{ww95}. The
model was expanded homologously and the gamma-ray deposition was
parameterized to be consistent with the nickel mixing found in
SN~1987A \citep{mitchetal87a01}. The abundances were taken directly from
the model, and the effects of radioactive decay were taken into
account. The results are summarized in Table~\ref{tab:dists}. The
explosion date is given as the number of days prior to discovery on
1999 October 29 (HJD 2451480.94). We used observed photometry of
\citet{leonard99em02} and \citet{hamuyepm01} in
$UBVRIZ$. The quoted errors are the $1-\sigma$ error in the
determination of the mean distance, which we believe are reasonably
accurate estimates of the true error which is difficult to determine
formally. For our favored value (see below) of $12.5$~Mpc we find a
formal error of 
$\pm 1.8$~Mpc if we add in quadrature the error in determining the
effective temperature ($\sim 500$~K), the error in determining the velocity
($\sim 500\ \kmps$), and the formal error in the mean.

\begin{deluxetable}{llll}
\tablecolumns{4}
\tablewidth{0pc}
\tablecaption{\label{tab:dists}}
\tablehead{
\colhead{Data Set}    &   \colhead{$\mu$}   &
\colhead{D (Mpc)} & \colhead{t$_{\mbox{exp}}$}}
\startdata
5 epochs including U & $30.07 \pm 0.8$ & $10.3 \pm 4.5$ & $5.2 \pm 0.4$\\
5 epochs excluding U & $30.47 \pm 0.39$ & $12.4 \pm 2.4$ & $5.9 \pm 0.3$\\
5 epochs excluding U\\
on 5th epoch &$30.49 \pm  0.36$ &$12.5 \pm 2.3$
&$5.9 \pm 0.3$\\
\enddata
\end{deluxetable}

\begin{figure}
\includegraphics[width=0.7\hsize,angle=90]{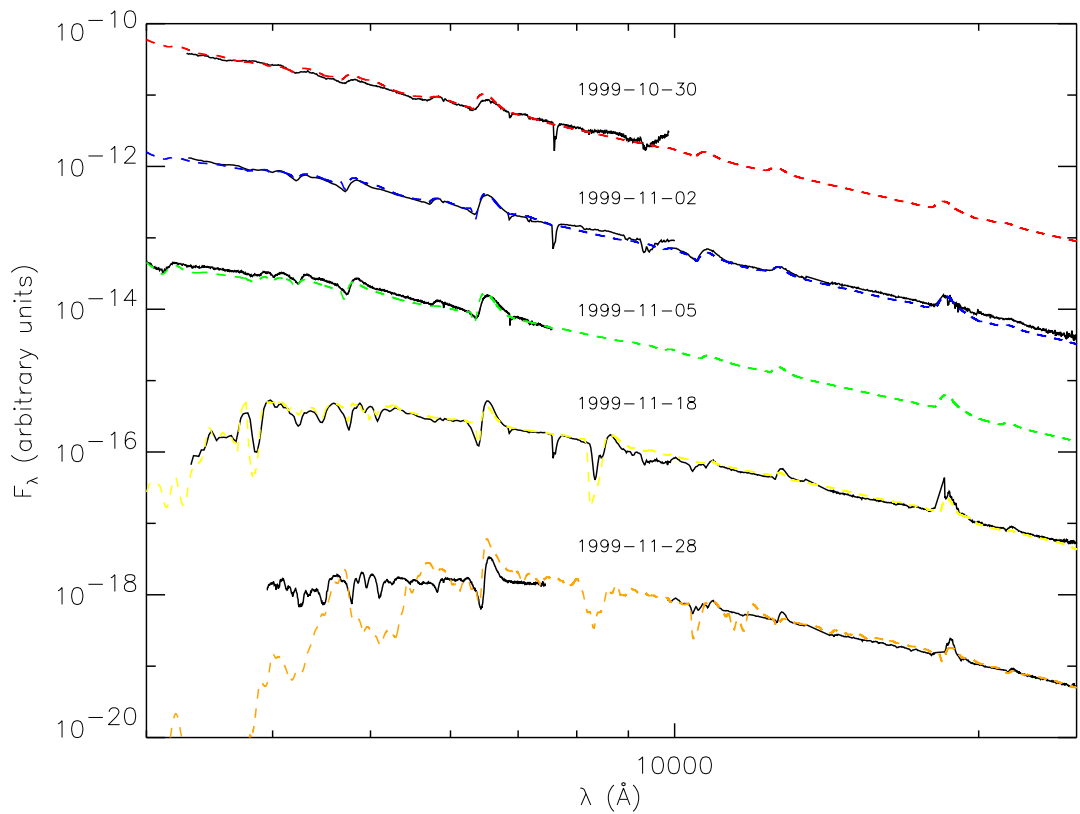}
\caption{\label{fig:fits}The synthetic spectra (dashed lines) are compared to
  observed spectra (solid lines) at 5 different epochs. The observed
  spectra were 
  obtained at CTIO for Oct 30, Nov 2, and Nov 18 \citep{hamuyepm01}, at
  \emph{HST} and FLWO on Nov 5 \citep{bsn99em00} and the optical
  spectrum on Nov 28 was obtained at Lick \citep{leonard99em02} while
  the IR was obtained at CTIO \citep{hamuyepm01}. The observed fluxes
  have been offset for clarity.}
\end{figure}

Figure~\ref{fig:fits} compares observed and model spectra,
details of the modeling will be discussed elsewhere. Overall the
fits are excellent, except on November 28 where the blue part of the
spectrum is poorly fit, this is due to the fact that at this late time
the spectrum forms over a much larger mass range of the ejecta and so we
are sensitive to the detailed mixing of both nickel and helium which
we have not attempted to adjust in the models. If we exclude the $U$
band from the calculation the scatter is considerably
reduced. Additionally, when the $U$ band is included the inferred
explosion date is nearer to the date of discovery which produces a
systematic rise in the SEAM distance with time. Errors in the
explosion date primarily affect the absolute magnitudes of the early
spectral models since they are more sensitive to errors in the
explosion date than are later epochs. If the estimated time from explosion is
too small, the models will have radii which are too small
($R=vt$). With smaller emitting area, they will be dimmer and hence
appear to be closer.The results of neglecting the $U$
band entirely are nearly identical with those if we include the $U$
data except for the one on November 28. The ability to compare 
synthetic spectra with observational spectra is
clearly an advantage of the SEAM method. Thus, we adopt the results of
the bottom line of Table~\ref{tab:dists}, which is in good agreement
with the Cepheid result and show that quality fits to SNe~IIP can give
distances accurate to 20\%, \emph{without adjusting metalicities,
  helium mixing, or nickel mixing}.  Once we have completed a large grid of
models which vary these parameters we should be able to reduce the
uncertainties even more, thus SNe~IIP will become important
cosmological probes.

\section{Discussion}

The SEAM method assumes that supernovae are spherically symmetric,
which is not strictly true. However, polarization data indicate that
SNe~IIP seem to be more spherically symmetric than other types of core
collapse supernovae, most likely because the large intact hydrogen
envelope sphericizes the explosion. Thus SNe~IIP appear to be the most
promising candidates for using the SEAM method. 
\citet{leonard01} found evidence for polarization in SN~1999em at
7--163 days after discovery. Modeled in terms of oblate electron
scattering atmospheres, the asphericity was about 7\%. They found some
tendency for increasing polarization with time.  This is consistent
with polarization studies of Type Ib/c supernovae where the
polarization appears to increase the closer one gets to the central
explosion mechanism \citep{wang02ap03}.

\begin{figure}
\includegraphics[width=0.7\hsize,angle=90]{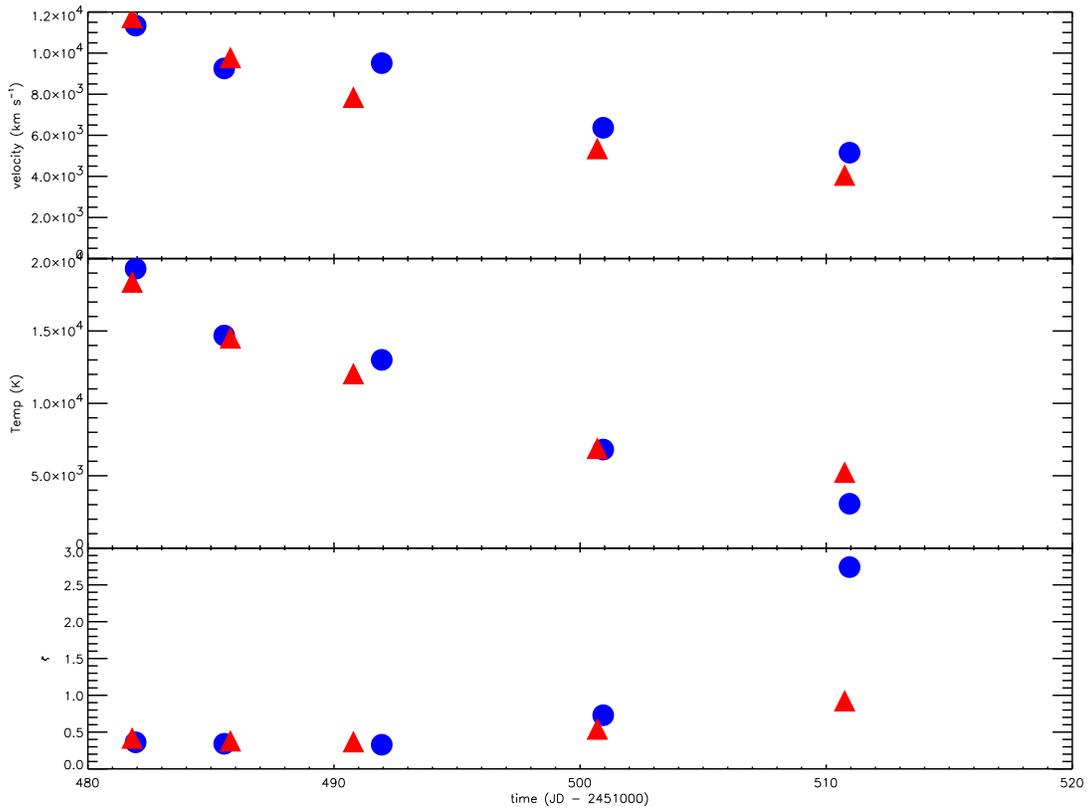}
\caption{\label{fig:epmcomp}The EPM parameters  $v(\tau
  =2/3)$, $T_{BV}$, and dilution factor $\zeta_{BV}$ from our models
  (filled circles) are 
  compared with those of \citet{hamuyepm01} (filled triangles). While
  there is good agreement at early epochs,
  by the fourth epoch the two results differ by 40\%.}
\end{figure}

\begin{figure}
\includegraphics[width=0.7\hsize,angle=90]{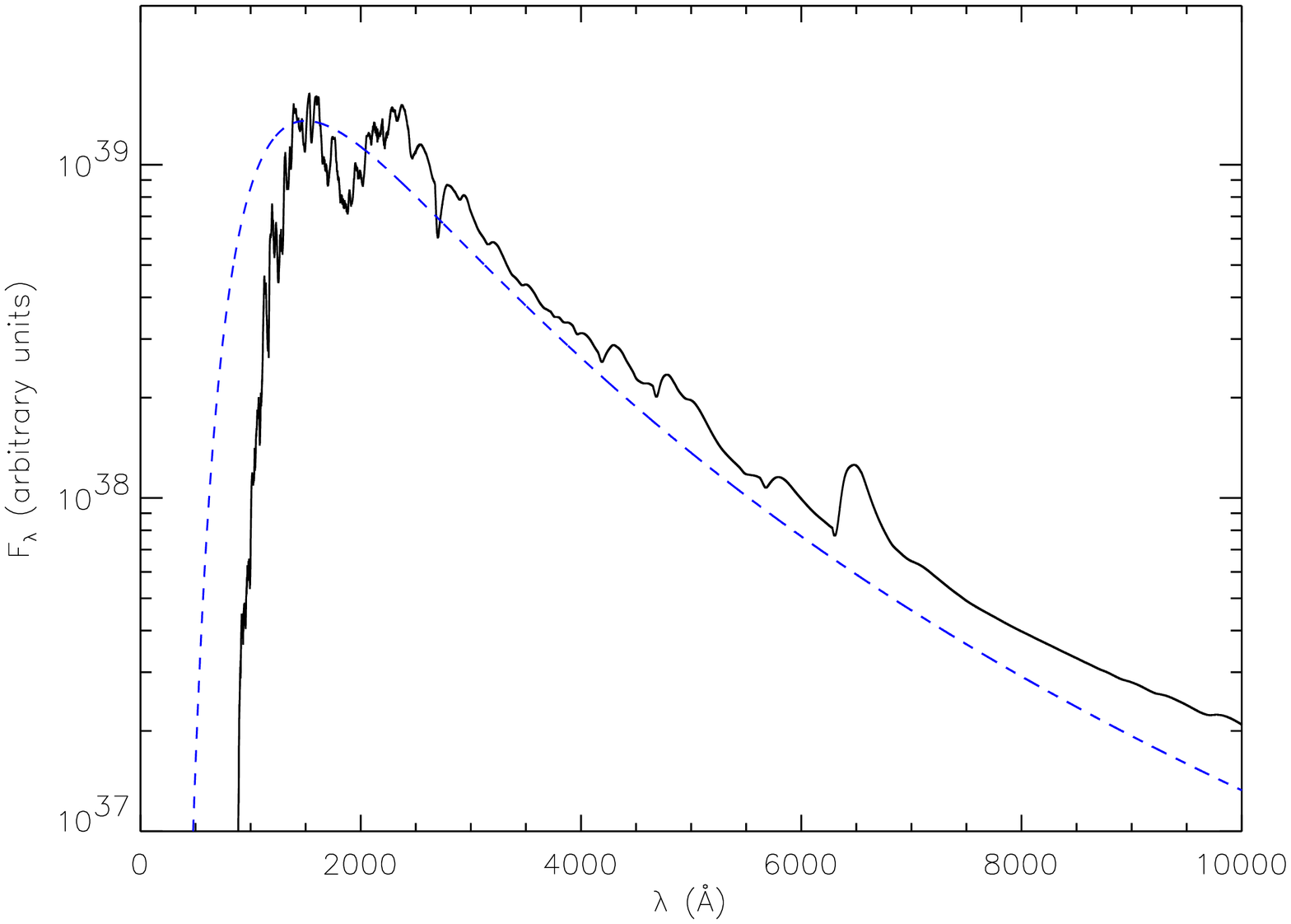}
\caption{\label{fig:bbcomp}The flux from our model (solid line)
  compared with the 
  best fit diluted blackbody flux (dashed line).}
\end{figure}

It is difficult to know exactly why the SEAM method gives such a
different result from that of EPM. \citet{leonard99em02}
found $t_{\mbox{exp}} = 5.3$~d, and our date is somewhat earlier.
Even with a similar explosion date (see Table~\ref{tab:dists}) we find
a larger distance. Figure~\ref{fig:epmcomp} compares the color
temperature $T_{BV}$, the velocity at the photosphere (defined as
$\tau =2/3$), and the dilution factor, $\zeta_{BV}$, obtained using
$T_{BV}$ with those of \citet{hamuyepm01}. The results agree
very well at early times, but by the 4th epoch the dilution factors
disagree by 40\% and by nearly a factor of 3 at the fifth
epoch. Comparing only two epochs, if 
one mistakenly uses a dilution factor that is too small at the later
time, the distance obtained will be too small. With hindsight
\citet{hamuyepm01} 
recognized this fact when they found that they obtained distances
close to the Cepheid value when they restricted their analysis to
early times where our dilution factors agree. However, the whole
foundation of EPM appears suspect. Figure~\ref{fig:bbcomp}
compares the best fit diluted Planck function with our computed flux
at the first epoch where we have fit the observations very well. It is
clear that a Planck function does not fit the SED at all.
Thus, we find that the
diluted blackbody assumption is too simplistic, particularly at later
times. That the EPM approach works at early times seems coincidental,
but it may be that in the hot early phases the color temperature is
reasonably accurate, we will explore this in detail in future work.

The SEAM method seems clearly superior to EPM since the assumption of
black-body emission is never realized in a supernova. SEAM should be
testable by the Nearby Supernova Factory \citep{snfactory02spie} if they
follow a dozen or so SNe IIP in the Hubble flow that they will
discover.  An independent cosmological probe is highly
desirable. 

SNe~IIP may be detectable to high redshifts with the James
Webb Space Telescope (\emph{JWST}). With a dataset of spectral models
that fit nearby SNe~IIP we will be able to determine the
nucleosynthetic history of the first generation of stars.

\begin{acknowledgments}
We thank Doug Leonard and Mario Hamuy for helpful discussions on SN
1999em and Type IIP supernovae. We thank the referee, Adam Riess for
improving the presentation of this work.
This work was supported in part by by NASA grant
NAG5-3505, NSF grants AST-0204771 and AST-0307323, an IBM SUR
grant to the University of Oklahoma and by 
NASA grants NAG 5-8425 and NAG 5-3619 to
the University of Georgia. PHH was 
supported in part by the P\^ole Scientifique de Mod\'elisation
Num\'erique at ENS-Lyon. 
 This research used resources
of: the San Diego Supercomputer Center (SDSC), supported by the NSF;
the National Energy Research Scientific Computing Center (NERSC),
which is supported by the Office of Science of the U.S.  Department of
Energy under Contract No. DE-AC03-76SF00098; and the
H\"ochstleistungs Rechenzentrum Nord (HLRN).  We thank all these
institutions for a generous allocation of computer time.
\end{acknowledgments}


\end{document}